\documentclass[fleqn,usenatbib]{mnras}

\usepackage{newtxtext,newtxmath}
\usepackage[T1]{fontenc}

\DeclareRobustCommand{\VAN}[3]{#2}
\let\VANthebibliography\thebibliography
\def\thebibliography{\DeclareRobustCommand{\VAN}[3]{##3}\VANthebibliography}

\usepackage{graphicx}	%
\usepackage{amsmath}	%

\title[C/O ratio effects on hot Jupiter interiors]{The effect of the atmospheric C/O ratio on interiors of hot Jupiters}

\author[van Dijk et al.]{
Esther van Dijk$^{1}$,\thanks{E-mail: evdijk@strw.leidenuniv.nl}
Yamila Miguel$^{1, 2}$,
Paul Mollière$^{3}$
\\
$^{1}$Leiden Observatory, Leiden University, Einsteinweg 55, 2333 CC, Leiden, the Netherlands\\
$^{2}$SRON Netherlands Institute for Space Research, Niels Bohrweg 4, 2333 CA, Leiden, the Netherlands\\
$^{3}$Max Planck Institute for Astronomy, Königstuhl 17, 69117 Heidelberg, Germany
}

\date{Accepted XXX. Received YYY; in original form ZZZ}

\pubyear{\the\year{}}

\begin{document}
\label{firstpage}
\pagerange{\pageref{firstpage}--\pageref{lastpage}}
\maketitle

\begin{abstract}

The atmosphere is the outer boundary of a gas giant exoplanet's interior. Therefore, compositional changes in the atmosphere can affect interior inferences. Typically, only metallicity is considered in atmospheric boundary conditions, while other elemental ratios, in particular the C/O ratio, are assumed to be solar. In light of the observational constraints now achievable with JWST, this assumption might no longer be justified.
In this work, we investigate the effect of the C/O ratio on atmospheric boundary conditions, interiors and radii of hot Jupiters.
We construct an atmospheric boundary grid, consisting of one-dimensional atmospheric models in radiative-convective and thermochemical equilibrium. This grid is coupled to a static interior structure model at the radiative-convective boundary (RCB).
Because in the temperature regime of hot Jupiters, the C/O ratio determines the dominant species in the atmosphere, it can significantly alter the opacity and, consequently, the pressure and temperature at the RCB. This temperature change propagates into the convective interior, thus affecting the calculated radius. The radius difference relative to a solar C/O-atmosphere significantly exceeds observed radius uncertainties of 3 per cent for planets with equilibrium temperatures $\gtrsim$ 1500 K and super-solar atmospheric metallicities. As an example, we demonstrate that, for WASP-19~b, higher C/O ratios result in higher inferred intrinsic temperatures.
These results highlight the importance of treating atmospheric composition and interior structure as a coupled system. As atmospheric constraints improve, incorporating measured C/O ratios into atmospheric boundary conditions is necessary to obtain robust inferences of hot Jupiter interiors.

\end{abstract}


\section{Introduction} 

Exoplanet interpretation requires a self-consistent understanding of both the atmosphere and interior. While spectral observations primarily probe the atmosphere, interior properties must be inferred indirectly from a set of limited observations constraints, such as the planetary mass, radius and age, combined with theoretical models \citep[e.g.][]{Thorngren_2016, Muller_2021, Muller_2023, Muller_2025, Acuna_2024, Acuna_2026, Chachan_2025, Thomas_2025, Peerani_2026}. Because the atmosphere provides the outer boundary condition for these models, its composition, opacity and temperature structure influence inferred interior properties, such as a planet's cooling history, radius, entropy and heavy element content. 

Several works have shown this dependency of interior properties on atmospheric characteristics. \citet{Poser_2024} showed that clouds significantly modify atmospheric opacity, influencing the cooling history and radius evolution of a planet. Including clouds can therefore introduce a degeneracy in the inferred bulk metallicity. Similarly, variations in atmospheric metallicity affect opacity and consequently the thermal evolution of a planet. \citet{Muller_2020}, \citet{Howard_2025} and \citet{Wilkinson2024} found that higher atmospheric metallicities lead to larger present-day radii, demonstrating that atmospheric metallicity can also influence the inferred interior properties.

However, metallicity alone does not uniquely determine atmospheric composition. In particular, the C/O ratio can change the dominant opacity sources in an atmosphere \citep[e.g.][]{Madhusudhan_2012, Molliere_2015, Goyal_2020}, because carbon and oxygen are the most abundant elements after hydrogen and helium in a solar composition atmosphere. Consequently, even at fixed metallicity, variations in C/O ratio may change atmospheric opacity and thermal structure with implications for interior properties and planetary evolution.

Despite the sensitivity of atmospheric chemistry to the C/O ratio, it is typically fixed at a solar value in atmospheric boundary conditions, reflecting the historical lack of observational constraints on this quantity. Because constraining the C/O ratio requires a broad wavelength range, moderate spectral resolution and high precision, such measurements have only recently become feasible with the launch of JWST \citep{JWST_mission}. With these data now becoming available \citep[e.g.][]{Ahrer_2023, Grant_2023_2, EvansSoma_2025, BOWIE_ALIGN_Kirk2025, BOWIE_ALIGN_Meech2025, BOWIE_ALIGN_Ahrer2025, BOWIE_ALIGN_Claringbold2026, BOWIE_ALIGN_Fairman2026}, ignoring C/O constraints might bias interior structure inferences.

Few studies have explored how the C/O ratio affects the interiors of gas giant exoplanets. \cite{Acuna_2024} found that for warm gas giants (T$_\mathrm{eq}$ $\leq$ 1000 K), the effect is only marginal. Their radius evolution predicts a 2\% difference between a C/O ratio of 0.2 and 0.55 at early evolutionary ages; this difference disappears at later ages. The effect on hot Jupiters remains unexplored, but may be stronger because of their higher atmospheric and intrinsic temperatures, leading to more extended atmospheres. 

Therefore, this study investigates the effect of the C/O ratio on the interiors of hot Jupiters. For that purpose, we develop an atmospheric boundary grid that, in addition to metallicity, also varies in C/O ratio. By coupling this grid to an interior model, we explore how metallicity and C/O ratio affect the atmospheric boundary condition and, consequently, the interiors of a range of synthetic planets. This approach allows us to identify the parameter regime in which the effect of the C/O ratio becomes significant and cannot be safely assumed to be solar. Lastly, the framework is applied to WASP-19\,b to quantify the impact of the C/O ratio on the inferred parameters given observational constraints.

\section{Interior-atmosphere model}

In this section, we present the coupled interior-atmosphere model, which expands on the model described in \cite{vandijk_2025} with new atmospheric boundary conditions, covering a range of metallicities, C/O ratios, intrinsic temperatures, effective temperatures and surface gravities. These boundary conditions are based on a grid of one-dimensional atmosphere models in radiative-convective thermochemical equilibrium, from which the associated boundary pressure and temperature are obtained by interpolation. The atmosphere model and the construction of atmospheric boundary conditions for the interior model are described in Section\,\ref{sec:atmosphere}. The interior model is presented in Section\,\ref{sec:interior} and the coupling between the interior model and the atmospheric grid is described in Section\,\ref{sec:coupling_interior_atmosphere}.

\subsection{Atmospheric boundary conditions} \label{sec:atmosphere}

\subsubsection{Atmosphere model} 

We generate a grid of 1D atmosphere models with the \textit{Pressure-Temperature Iterator and Spectral Emission Calculator for Planetary Atmospheres} (\texttt{petitCODE}) \citep{Molliere_2015, Molliere_2017}. This code computes a pressure-temperature profile in radiative-convective and thermochemical equilibrium by iterating between chemistry calculations and radiative transfer calculations until convergence. To obtain radiative-convective equilibrium, the Schwarzschild criterion is evaluated at each atmospheric layer throughout the iterative procedure.

Within this framework, chemical equilibrium is computed using \texttt{easyCHEM} \citep{Molliere_2017}. The chemical network includes, apart from abundant gas species, several condensates, such as VO, MgSiO$_3$, Mg$_2$SiO$_4$, SiC, Fe, Al$_2$O$_3$, Na$_2$S, KCl, H$_2$O, TiO, MgAl$_2$O$_4$, FeO, Fe$_2$O$_3$, Fe$_2$SiO$_4$, TiO$_2$ and H$_3$PO$_4$. We assume a solar elemental composition \citep{Asplund_2021}, which can be adapted by setting a metallicity and a C/O ratio. The metallicity is varied by scaling the abundances of all elements heavier than helium by a constant enrichment factor and the C/O ratio is modified by adjusting both the carbon and oxygen abundance relative to hydrogen, while conserving $(N_{\rm O} + N_{\rm C})/N_{\rm H}$. 

Chemical equilibrium calculations are alternated with radiative-transfer calculations, which are performed with the correlated-k method. Before the iterative procedure, \texttt{petitCODE} pre-computes a correlated-k opacity table on a grid in pressure-temperature space. Then, throughout the iterative procedure, opacities are obtained by interpolation within this grid, reducing computational cost. The choice of opacity grid spacing (40x40) and spectral resolution ( $\lambda/\Delta \lambda = 10$) is sufficient to ensure accurate results \citep{Molliere_2015}. 

We include a broad set of opacity sources. Molecular line opacities are included for CH$_4$, H$_2$O, CO$_2$, HCN, CO, H$_2$, H$_2$S, NH$_3$, OH, C$_2$H$_2$, PH$_3$, TiO, VO, and atomic Na, K, Fe and Mg. Continuum opacities include collision induced absorption (CIA) opacities from H$_2$-H$_2$ and H$_2$-He, as well as H$^-$ bound-free and free-free absorption. Rayleigh scattering is included for H$_2$, He, H$_2$O, CO$_2$, CO and CH$_4$. Line lists for H$_2$O, CO$_2$, H$_2$S, NH$_3$, OH, C$_2$H$_2$ and PH$_3$ are taken from HITRAN/HITEMP databases \citep{Rothman2013, Rothman2010}. Additional sources include Kurucz line lists for CO, H$_2$, Fe, Mg \citep{Kurucz}, ExoMol data for HCN \citep{Harris_2006, Barber_2014}, and atomic line data for K, Na \citep{Piskunov}. TiO and VO are provided by B. Plez (priv. comm.). For CH$_4$, we use HITRAN for temperatures below 300 K and ExoMol cross-sections at higher temperatures \citep{Yurchenko_2014}.  

The incident stellar radiation is represented by a solar-like spectral energy distribution with a stellar effective temperature of $T_* = 5772$ K. We parametrize the incident stellar flux in terms of the irradiation temperature $T_{\rm irr} = T_* \sqrt{\frac{R_*}{a}},$ where $R_*$ is the stellar radius and $a$ is the semi-major axis. For a given irradiation strength, the corresponding zero-albedo equilibrium temperature depends on the assumed redistribution of the incident stellar flux. For a planetary-average atmosphere, $T_{\rm eq,0}^{\rm p} = 4^{-1/4}T_{\rm irr}$, whereas for a dayside-average atmosphere, $T_{\rm eq,0}^{\rm d} = 2^{-1/4}T_{\rm irr}$. The subscript 0 denotes that these temperatures describe the incident stellar flux before accounting for reflection. The planetary effective temperature is used to parametrize the model grid and is related to the internal and irradiation fluxes through $T_{\rm eff}^4 = T_{\rm int}^4 + \left(T_{\rm eq,0}^{\rm p}\right)^4$. Thus, by specifying $T_{\rm eff}$ and $T_{\rm int}$ \texttt{petitCODE} determines the incident stellar flux required for the converged model. 

The grid is calculated assuming planetary-average irradiation. For hot Jupiters, the observed thermal emission predominantly probes the dayside, for which a dayside-average irradiation treatment may be more appropriate than the planetary-average assumption used to construct the grid. The equivalent dayside irradiation level can be represented using the planetary-average grid by converting between the corresponding equilibrium temperatures, $T_{\rm eq,0}^{\rm d} = 2^{1/4}T_{\rm eq,0}^{\rm p}$. Hereafter, we denote $T_{\rm eq,0}^{\rm p}$ simply as the equilibrium temperature $T_{\rm eq}$.

\texttt{petitCODE} treats the stellar and planetary radiation fields separately within the radiative-transfer calculation. The stellar component accounts for absorption and (multiple) scattering of the incident stellar radiation, while planetary thermal emission is treated separately. The upward stellar flux emerging from the top of the atmosphere therefore corresponds to the stellar radiation reflected by the atmosphere. The Bond albedo is calculated self-consistently by integrating the reflected stellar flux over wavelength and angle and taking the ratio of the total reflected stellar flux to the incident stellar flux. Consequently, the Bond albedo is not prescribed as an input parameter, but follows from the atmospheric structure, wavelength-dependent opacities, and scattering properties during the iterative radiative-convective equilibrium calculation.

The atmosphere is discretized into 130 layers spanning pressures from $10^{-7}$ bar to $10^{3}$ bar. For some models with high intrinsic temperatures, high metallicity and/or low surface gravity, the lower boundary is adjusted to lower pressures due to the upper temperature limit of \texttt{easyCHEM} (20,000 K). This prevents numerical instabilities in the code that may occur before convection is turned on during the iterative solution process.

\subsubsection{Convergence} \label{sec:convergence}
A model is considered converged when two criteria are simultaneously satisfied: (i) the maximum change in temperature between the current iteration and 60 iterations prior is less than 0.01 K, and (ii) the relative difference between the emergent flux at the top of the atmosphere and the imposed flux is below 0.001.

In some cases, these convergence criteria are not met. We divide such non-converged models into two categories. The first category comprises models that achieve temperature convergence in the deep atmosphere, typically for pressures higher than 10 mbar, but fail to converge in the upper atmosphere. These models remain suitable as boundary conditions for interior structure and evolution calculations.

The second category includes 18 models that fail to converge due to ambiguity in the dominant energy transport mechanism (radiative versus convective) across multiple layers in the upper atmosphere. This issue is a known limitation of \texttt{petitCODE}, particularly for atmospheres with high metallicity and/or high temperatures \citep{Molliere_2015}. For these cases, we recompute the models by enforcing radiative equilibrium in all layers at pressures below 10 mbar. We emphasize that these modified models should not be used for upper-atmosphere analyses. However, since the focus of this work is the deep atmosphere and interior and radiative-convective equilibrium is still calculated for these layers, these solutions remain sufficient for our purposes.

The models that remain suitable for interior structure calculations, but should not be used for upper-atmosphere analyses are flagged with '\_false' in the filename in the accompanying Zenodo repository \footnote{\url{https://doi.org/10.5281/zenodo.20629944}}.

\subsubsection{Grid values and interpolation} \label{sec:grid_interpolation}
The computed atmospheric grid spans a range in metallicity (-2.0 $\leq$ $[M/H]$ $\leq$ 2.0 ), C/O ratio (0.2 $\leq$ $C/O$ $\leq$ 1.4), intrinsic temperature (70 K $\leq$ $T_\mathrm{int}$ $\leq$ 1050 K), effective temperature (1000 K $\leq$ $T_\mathrm{eff}$ $\leq$ 3000 K) and surface gravity (2.3 $\leq$ log $g$ $\leq$ 5.0). The discrete grid points are listed in Table \ref{tab:grid_points_atmosphere}. 

For every atmospheric model in the grid, we save the full atmospheric temperature, pressure and density profiles. In addition, we calculate the mass fractions of hydrogen, helium and metals as a function of pressure. These mass fractions are calculated by summing the mass mixing ratios, computed under the assumption of equilibrium chemistry (with \texttt{easyCHEM}), of the corresponding molecular and atomic species. For hydrogen, these include H, H$_2$, H$^+$ and H$^-$; for helium, only He; and the metal mass fraction comprises all remaining species. Lastly, the location of the RCB is saved for every atmospheric grid point.

The atmospheric model is coupled to the interior model at the RCB (see Section\,\ref{sec:coupling_interior_atmosphere}). Therefore, the outer boundary condition consists of the pressure, temperature and composition (hydrogen, helium and metal mass fractions, that together sum up to 1) at the RCB. To determine boundary conditions for a given atmospheric configuration, we linearly interpolate the atmospheric profiles using Python's \texttt{scipy.interpolate.RegularGridInterpolator} to obtain the P-T-$\rho$-X-Y-Z profiles. The RCB pressure is interpolated separately. By combining this interpolated RCB pressure with the interpolated atmospheric profiles, we determine the pressure, temperature and mass fractions at the RCB. 

To evaluate whether the grid resolution is sufficient, we generate a set of randomly selected coupled interior–atmosphere models and recompute their atmospheric structures using \texttt{petitCODE}. These non-interpolated atmospheric profiles are used as direct input for the interior models and compared to the original model output. The resulting differences in the planetary radius are typically below 2\%, which is less than typical observational uncertainties.

\begin{table}
    \centering
    \caption{Grid points used to create the atmospheric boundary conditions}
    \begin{tabular}{l|c}
    \hline
    Parameter & Grid values \\ \hline
     $[M/H]$    &  -2.0, -0.5, 0.0, 0.5, 1.0, 1.5, 2.0 \\
     $C/O$    & 0.2, 0.53, 0.95, 1.4 \\
     $T_\mathrm{int}$ [K]  & 70, 100, 150, 250, 350, 450, 550, 650, 750, 850, 950, 1050 \\
     $T_\mathrm{eff}$ [K]  & 1000, 1500, 2000, 2500, 3000 \\
     log $g$ [cm/s$^2$]  & 2.3, 3.0, 4.0, 5.0\\
     \hline
    \end{tabular}   
    \label{tab:grid_points_atmosphere}
\end{table}

\subsection{Interior} \label{sec:interior}

The interior structure is modelled with \texttt{CEPAM} \citep{Guillot1995}. \texttt{CEPAM} iteratively solves the interior structure equations, for which it needs two boundary conditions. The bottom boundary is set by a rocky isothermal core, approximated with an analytical expression \citep{Hubbard_1989}. The upper boundary is the new atmospheric grid. In between the two boundary conditions, there is a homogeneous envelope, which can contain hydrogen, helium and heavy elements. The equations of state used are \cite{Chabrier_2019} for hydrogen and helium with non-ideal mixing coefficients of \cite{Howard_2023}. We represent metals with water \citep{Mazevet_2019}. 

Usually, the interior is fully convective, as we set the atmospheric boundary limit at the RCB. However, in some cases, the RCB can be at pressures higher than the maximum depth of the atmosphere ($10^3$ bar). In that case, \texttt{CEPAM} calculates the Schwarzschild criterium to divide the interior into radiative and convective regions. For this purpose, the code uses the Rosseland mean opacities calculated by \cite{Freedman_2008} and a fit that accounts for different metallicities by \cite{Valencia_2013}. 

The parameters that describe the interior are the core mass fraction ($m_\mathrm{core}$); the envelope hydrogen ($X_\mathrm{env}$), helium ($Y_\mathrm{env}$) and metal mass fraction ($Z_\mathrm{env}$) (where $X_{\rm env} + Y_{\rm env} + Z_{\rm env} = 1$); the mass of the planet ($M_\mathrm{planet}$) and the intrinsic luminosity ($L_\mathrm{int}$). The intrinsic luminosity parametrizes the outgoing flux from the interior and is defined by $L_{\rm int} = 4\pi \sigma R_p^2 (T_{\rm eff}^4 - T_{\rm eq}^4)$. In the case of a fully convective interior the intrinsic luminosity is only used to calculate the atmospheric boundary (see Section\,\ref{sec:coupling_interior_atmosphere}) and is not needed in the interior model calculation. However, in the above-mentioned case of a not-fully convective interior, the intrinsic luminosity is required to calculate the radiative gradient.

\subsection{Coupling between interior and atmosphere} \label{sec:coupling_interior_atmosphere}

\begin{figure}
    \centering
    \includegraphics[width=\columnwidth]{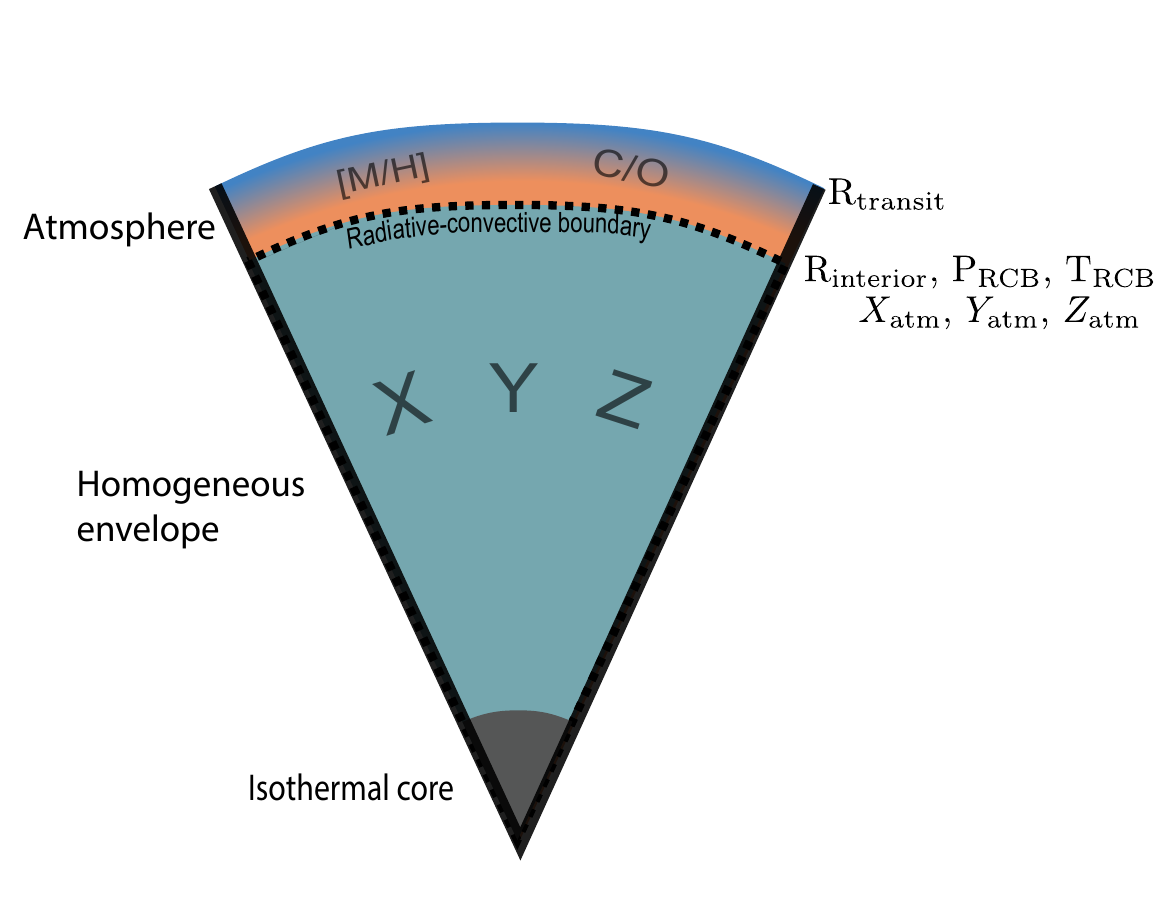}
    \caption{Schematic illustration of the coupled interior-atmosphere model. The interior has an isothermal core and a homogeneous envelope with hydrogen, helium and metals. The composition of the envelope is set through the mass fractions $X$ (hydrogen), $Y$ (helium) and $Z$ (metals). The atmosphere starts at the radiative-convective boundary, where P$_\mathrm{RCB}$ and T$_\mathrm{RCB}$ provide the outer boundary condition for the interior. The atmospheric composition at the RCB, also defined by mass fractions $X_\mathrm{atm}$, $Y_\mathrm{atm}$ and $Z_\mathrm{atm}$, is used to constrain the envelope composition. The full radius is calculated by summing the radius at the RCB (R$_\mathrm{interior}$) and the atmospheric contribution to the radius. We define the transit radius (R$_\mathrm{transit}$) at a pressure of 10 mbar.}
    \label{fig:coupling_interior_atmosphere}
\end{figure}

A schematic illustration of the coupled interior-atmosphere model is shown in Fig.~\ref{fig:coupling_interior_atmosphere}. We couple the interior and atmosphere model at either the RCB or at a pressure of 1 kbar if the RCB lies deeper than this pressure, which may occur for low intrinsic temperatures, high effective temperatures, high surface gravities and/or low metallicities. 

The coupled interior-atmosphere model is parametrized by the planetary mass ($M_\mathrm{planet}$), the core mass fraction ($m_\mathrm{core}$), the intrinsic luminosity ($L_\mathrm{int}$), the equilibrium temperature ($T_\mathrm{eq}$) and the atmospheric composition (metallicity and C/O ratio). To determine the atmospheric boundary conditions required by the interior model, these parameters are first mapped onto the corresponding inputs of the atmosphere model. The intrinsic temperature is calculated from the intrinsic luminosity with the Stefan-Boltzmann law ($T_\mathrm{int} = (L_\mathrm{int} / (4 \pi R_p^2))^{1/4}$, the surface gravity is calculated from the planetary mass ($g = M_p/R_p^2$) and the effective temperature is calculated with $T_\mathrm{eff} = (T_\mathrm{eq}^4 + T_\mathrm{int}^4)^{1/4}$. These calculations require a planetary radius, but, because the radius is an output of the model, this is not known a priori. Therefore, we perform an iterative procedure, in which we adopt a guess ($R_\mathrm{guess}$) for the transit radius ($R_\mathrm{transit}$) of the planet for the first iteration.

We interpolate the atmospheric boundary grid to get the corresponding pressure ($P_\mathrm{RCB}$), temperature ($T_\mathrm{RCB}$) and mass fractions of hydrogen ($X_\mathrm{atm}$), helium ($Y_\mathrm{atm}$) and metals ($Z_\mathrm{atm}$) at the RCB, as described in Section \ref{sec:grid_interpolation}. The pressure and temperature are used as an outer boundary condition for CEPAM and the mass fractions at the RCB set the envelope composition in the interior ($X_\mathrm{env}=X_\mathrm{atm}$, $Y_\mathrm{env}=Y_\mathrm{atm}$, $Z_\mathrm{atm}=Z_\mathrm{env}$). Together with the additional inputs for the interior ($m_\mathrm{core}$, $M_\mathrm{planet}$ and $L_\mathrm{int}$), \texttt{CEPAM} calculates the radius at the RCB (R$_\mathrm{interior}$). 

The radius at the RCB does not represent the observed transit radius yet, as the contribution from the atmosphere is missing. We define the transit radius at a pressure of 10 mbar. In reality, this depends on the wavelength and the opacities in the atmosphere. In Section\,\ref{sec:transit_radius_definition}, we discuss this approximation. The contribution to the radius from the atmosphere is calculated with

\begin{equation}
    \Delta R_\mathrm{atm} = - \sum_{i=(P=P_\mathrm{RCB})}^{P=\mathrm{10 mbar}} \frac{ \delta P_i}{\rho_i g},
\end{equation}

where $\delta P_i$ is the pressure difference between subsequent layers, $\rho_i$ is the density in layer $i$ and $g$ is the surface gravity. The densities for pressures greater than 1 bar are recalculated using the equations of state specified in Section \ref{sec:interior}. For pressures lower than 1 bar, we use the densities directly from the atmospheric model (calculated assuming an ideal gas equation of state).

Every subsequent iteration, $R_\mathrm{guess}$ is replaced with the calculated $R_\mathrm{transit}$. We iterate between interior and atmosphere until the difference between $R_\mathrm{guess}$ and $R_\mathrm{transit}$ is 0.1\%. 

Apart from radius, another observable that can be measured for some hot Jupiters is the Love number. For example, our case-study WASP-19\,b (see Section\,\ref{sec:WASP19b_retrieval}) has one of the most precise measurements to date \citep{Bernabo_2024}. We calculate the Love number from the full density profile (interior + atmosphere), so that the Love number is defined by the Love function at the transit radius ($k_2 = K_2(R_\mathrm{transit})$) \citep{Love}. The Love number can be calculated from the density profile by solving equations (4) and (5) in  \citet{vandijk_2025}. This calculation provides the static Love number to leading order for fluid planets, neglecting nonlinearities in the response. Nonlinear contributions may arise for rapidly rotating planets, but can safely be ignored for planets with $q_0 \ll 0.01$, where $q_0$ is the ratio of centrifugal to gravitational acceleration \citep{Wahl_2021}. Additionally, this approach neglects dynamical contributions arising from, e.g., departures from tidal synchronisation, non-zero eccentricity or non-zero obliquity. These approximations are generally valid for hot Jupiters that are tidally locked and have circularized orbits.

\section{Synthetic planets}

We investigate the effect of atmospheric composition on the planetary interior using synthetic planets. Each synthetic planet is defined by a fixed planet mass, core mass fraction and equilibrium temperature, whereas the atmospheric composition is varied. Thereby, we isolate the impact of atmospheric composition on the outer boundary condition and, as a result, on the calculated radius. We consider a 1 $M_\mathrm{J}$ planet with a core mass fraction of 0.1 and an equilibrium temperature of 2500 K. The intrinsic luminosity, which is generally unknown for hot Jupiters, is varied across simulations.

\subsection{Effect of atmospheric metallicity} \label{sec:effect_of_metallicity}

\begin{figure}
    \centering
    \includegraphics[width=\columnwidth]{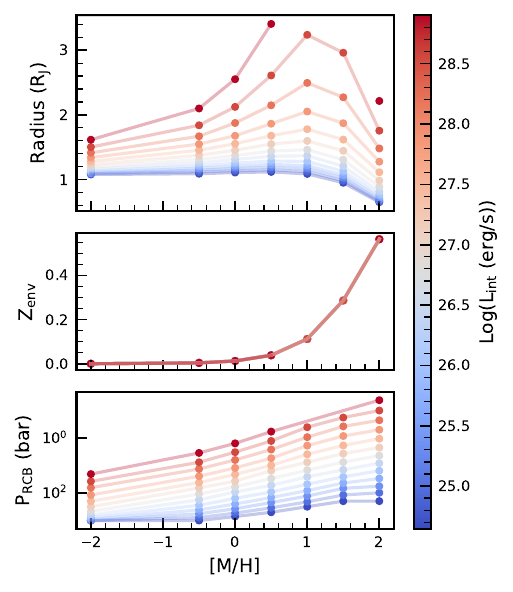}
    \caption{Variation of radius, envelope metal mass fraction and RCB pressure as a function of atmospheric metallicity for different intrinsic luminosities. We use a 1 M$_\mathrm{J}$ planet with an equilibrium temperature of 2500 K, a core mass fraction of 0.1 and an atmospheric C/O ratio of 0.55.}
    \label{fig:effect_of_metallicity_oneplanet}
\end{figure}

\begin{figure*}
    \centering
    \includegraphics[width=1.2\columnwidth]{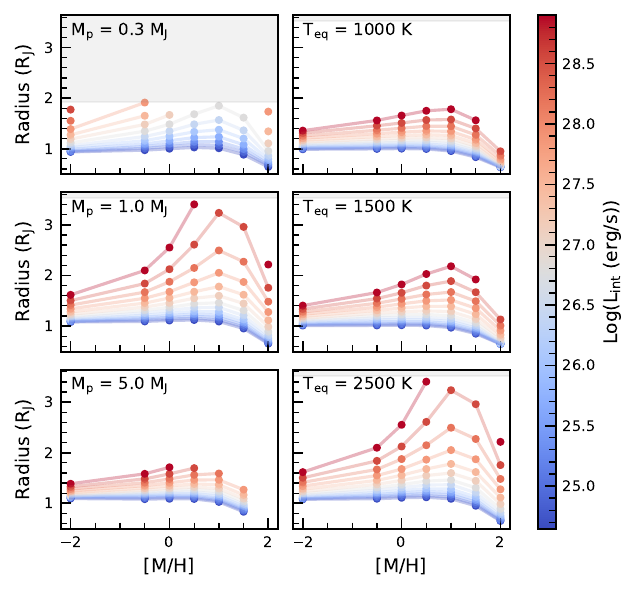}
    \caption{Variation in radius with atmospheric metallicity for different planetary masses and equilibrium temperatures. For all simulations, the core mass is fixed to 0.1 $M_\mathrm{J}$ and the atmospheric C/O ratio to 0.55. The first column shows results for different planetary masses. Here, the equilibrium temperature is fixed to 2500 K. The second column shows results for different equilibrium temperatures, while fixing the planetary mass to 1 $M_\mathrm{J}$. The grey shaded areas indicate radii outside the atmospheric grid with log $g$ > 5.0 or log $g$ < 2.3.}
    \label{fig:effect_of_metallicity_differentvariations}
\end{figure*}

Fig.~\ref{fig:effect_of_metallicity_oneplanet} shows the variation in radius, envelope metal mass fraction and RCB pressure with atmospheric metallicity for the synthetic planet. The C/O ratio is fixed to 0.55. The envelope metal fraction in our models, defined as the metal mass fraction at the RCB in the atmosphere, is directly coupled to the atmospheric metallicity through equilibrium chemistry. The found metallicity-metal mass fraction trend (middle panel) is similar to other conversions used in the field (see e.g. appendix\,A of \citet{Fortney_2013}), with some small deviations that we explain in Appendix\,\ref{ap:metallicity_conversion}. The metal mass fraction increases with higher atmospheric metallicity and consequently causes a higher molecular weight in the envelope. This effect in isolation would solely lead to a decrease in radius. However, at the same time the RCB moves to lower pressures for higher atmospheric metallicity due to higher opacities in the atmosphere (bottom panel). For the interior, this means that the adiabat starts at a lower pressure, resulting in hotter temperatures in the interior, which inflates the radius. 

For low intrinsic luminosity (blue), the envelope metal fraction plays the largest role. The radius is almost constant for low metallicity and starts decreasing above a metallicity of 3 $\times$ solar. For higher intrinsic luminosities (red), the radius first increases due to the increasing opacities with metallicity until a metallicity of approximately 10 $\times$ solar, after which the radius decreases due to higher envelope metallicities and consequent increase in mean molecular weight. The increase in radius due to higher opacities is greater at higher intrinsic luminosities for two main reasons. Firstly, for higher intrinsic luminosities, the RCB moves faster to lower pressures when the atmospheric metallicity increases, especially at high metallicity.  Secondly, planets with higher intrinsic luminosities are more inflated and have larger scale heights. A change in temperature will therefore induce a larger radius difference than for less inflated planets. The trend of radius with atmospheric metallicity that we find is consistent with previous studies \citep[e.g.][]{Wilkinson2024}. 

In Fig.~\ref{fig:effect_of_metallicity_differentvariations}, we show how the behaviour changes for different planetary masses and equilibrium temperatures. We generally find that radius differences become larger for lower planetary masses and higher equilibrium temperatures, because the scale height in the atmosphere increases with both lower gravity (lower planetary masses) and higher temperatures (higher equilibrium temperatures).

\subsection{Effect of C/O ratio} \label{sec:effect_of_coratio}

\begin{figure}
    \centering
    \includegraphics[width=\columnwidth]{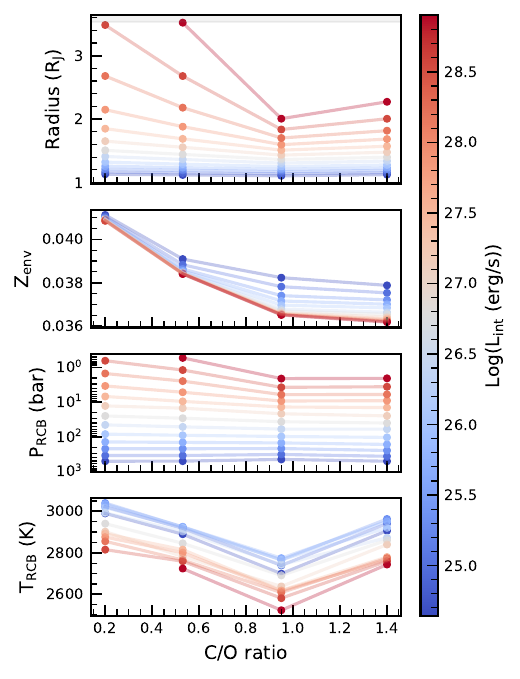}
    \caption{Variation of radius, envelope metal mass fraction, RCB pressure and RCB temperature as a function of atmospheric C/O ratio for different intrinsic luminosities. We use a 1 M$_\mathrm{J}$ planet with an equilibrium temperature of 2500 K, a core mass fraction of 0.1 and an atmospheric metallicity of 3 $\times$ solar.}
    \label{fig:effect_of_co_oneplanet}
\end{figure}

\begin{figure}
    \centering
    \includegraphics[width=\columnwidth]{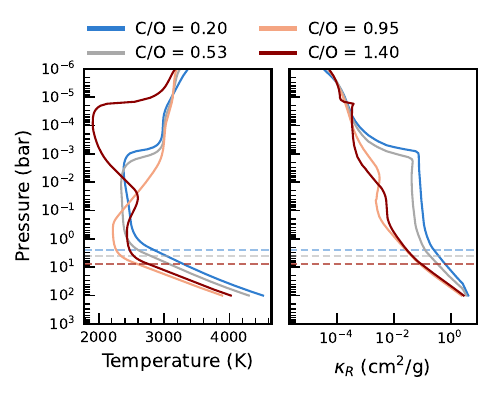}
    \caption{Atmospheric pressure-temperature profiles and Rosseland mean opacities for different C/O ratios for T$_\mathrm{eff}$ = 2500 K, T$_\mathrm{int}$ = 550 K, [M/H] = 0.5, log $g$ = 3.0.}
    \label{fig:PT_profiles_different_COratios}
\end{figure}

\begin{figure*}
    \centering
    \includegraphics[width=1.5\columnwidth]{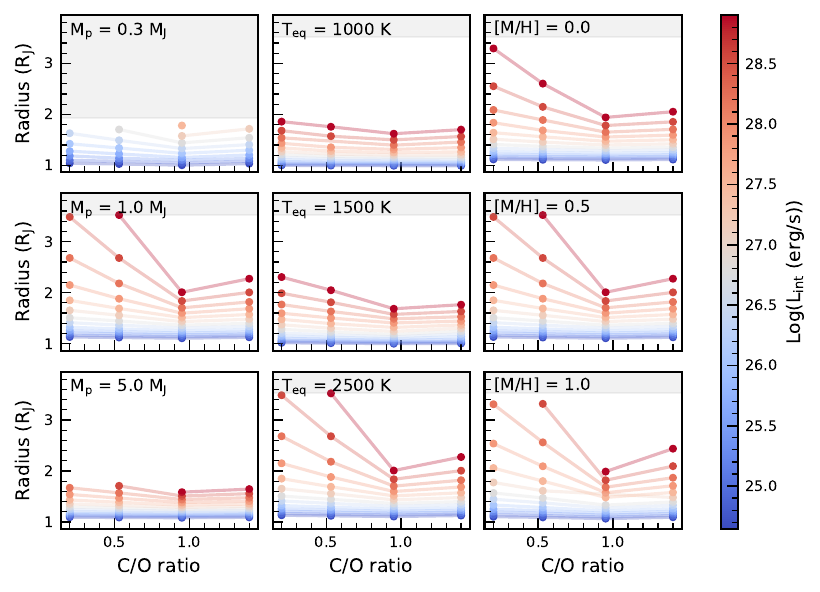}
    \caption{Variation in radius with C/O ratio for different planetary masses, equilibrium temperatures and atmospheric metallicities. In the first column, the planetary mass is varied for a planet with an equilibrium temperature of 2500 K and an atmospheric metallicity of 3 $\times$ solar. In the second column, the equilibrium temperature is varied for a 1 M$_\mathrm{J}$ planet with an atmospheric metallicity of 3 $\times$ solar. In the third column, the atmospheric metallicity is varied for a 1 M$_\mathrm{J}$ planet with an equilibrium temperature of 2500 K. The grey shaded areas indicate radii outside the atmospheric grid with log $g$ > 5.0 or log $g$ < 2.3.}
    \label{fig:effect_of_co_differentvariations}
\end{figure*}

Fig.~\ref{fig:effect_of_co_oneplanet} shows the variation of radius, envelope metal mass fraction, RCB pressure and RCB temperature as a function of atmospheric C/O ratio for the synthetic planet. The atmospheric metallicity is fixed to 3 $\times$ solar. The effect of the C/O ratio on the atmospheric boundary is different for a C/O ratio smaller and larger than approximately 1, which can be explained by the different chemical regimes before and after this turning point. 

At high atmospheric temperatures (>1300\,K at 1\,bar), the conversion of CH$_4$ to CO is favoured \citep[e.g.][]{Burrows_Sharp_1999, Goukenleuque_2000, Seager_2000, Sudarsky_2003, Madhusudhan_2012}. As a result, the CO abundance for different C/O ratios is approximately constant. The abundances of other species are therefore dependent on the amount of excess oxygen and carbon, which is determined by the C/O ratio.  At the turning point (C/O$\approx$1), where the composition changes from an oxygen- to a carbon-dominated atmosphere \citep[e.g.][]{Madhusudhan_2012, Moses_2013}, there is little excess oxygen and carbon, resulting in a depletion of major opacity carriers (e.g. H$_2$O, CH$_4$ and C$_2$H$_2$), which causes a minimum in opacity. Therefore, at C/O=0.95, both the RCB temperature and radius are the lowest.

At C/O<1, excess oxygen forms predominantly H$_2$O. Its abundance increases with lower C/O ratio, causing an increase in Rosseland mean opacities (see the right panel of Fig.~\ref{fig:PT_profiles_different_COratios}). This increase in opacity moves the RCB to lower pressures and increases the temperature at the RCB by 200-300~K, depending on the intrinsic luminosity (see Fig.~\ref{fig:effect_of_co_oneplanet}, rows 3-4). Together, these effects lead to higher interior temperatures for lower C/O ratios, resulting in a larger radius (see Fig.~\ref{fig:effect_of_co_oneplanet}, row 1). 

At C/O>1, the H$_2$O abundance is low, while carbon-rich species such as CH$_4$, HCN, and C$_2$H$_2$ become more abundant. The higher abundance of carbon-rich molecules, especially C$_2$H$_2$ and HCN, increases the atmospheric opacity, raising the temperature at the RCB compared to the C/O=0.95 case. Additionally, the depletion of TiO and VO at high C/O cause a weaker thermal inversion in the temperature structure, shown in Fig.~\ref{fig:PT_profiles_different_COratios}, which may also influence the temperature at the RCB. The higher temperatures at the RCB (by 100-200 K, depending on the intrinsic luminosity) lead, also in this case, to higher overall temperatures in the interior and a larger radius with higher C/O ratio. 

The exact C/O ratio where the transition from an O-dominated to a C-dominated atmosphere occurs depends on the temperature and can range from 0.73 to 0.92 \citep{Molliere_2015}. We do not resolve the exact transition location due to the coarse sampling of C/O in the atmospheric grid. However, we are able to resolve the general trend by including one grid point close to the transition at C/O = 0.95. 

The envelope metallicity changes marginally for different C/O ratios (from $\sim 0.041$ for a C/O ratio of 0.2 to $\sim0.036$ for a C/O ratio of 1.4 and the highest intrinsic luminosity modelled). This change follows from our adopted definitions of metallicity and C/O ratio and the changing mass-budget of hydrogen locked up in metal-molecules for different C/O ratios (see Appendix\,\ref{ap:metallicity_conversion}). Our adopted definitions of metallicity and C/O ratio are constructed such that they conserve the total number density of carbon and oxygen. Because oxygen is heavier than carbon, a higher abundance of oxygen increases the total C+O mass, resulting in a higher metal mass fraction with lower C/O ratio. However, this compositional difference ($\sim 0.041$ to $\sim0.036$) is negligible, as the corresponding radius difference is only 0.14 per cent for a 1\,$M_{\rm J}$ planet with a core mass fraction of 0.1, an atmospheric metallicity of 0.5 dex, a C/O ratio of 0.2 and an intrinsic luminosity of $3\cdot 10^{28}$\,erg\,s$^{-1}$.

In Fig.~\ref{fig:effect_of_co_differentvariations}, we show the variation in radius with C/O ratio for different masses (left), different equilibrium temperatures (middle) and atmospheric metallicities (right). The radius response is stronger for lower-mass planets and higher equilibrium temperatures, consistent with their larger atmospheric scale heights (as also seen in Section~\ref{sec:effect_of_metallicity}). In addition, increasing atmospheric metallicity also amplifies the effect, because there is a higher abundance of opacity-driving molecules (H$_2$O, CH$_4$, C$_2$H$_2$) that increase the opacity and thus raise temperatures in the atmosphere at both C/O<1 and C/O>1.

\subsubsection{Validity of the solar C/O assumption} \label{sec:comparison_solarCO}

\begin{figure*}
    \centering
    \includegraphics[width=2\columnwidth]{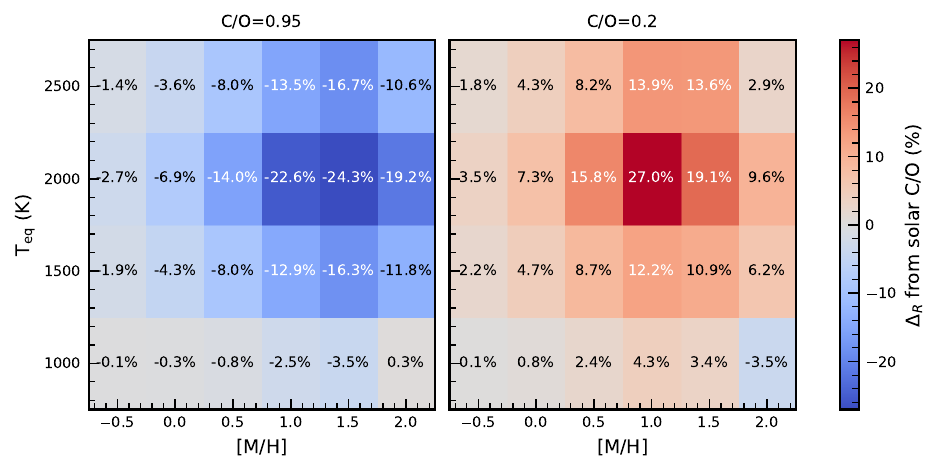}
    \caption{Radius difference between a C/O ratio of 0.95 (left) / 0.2 (right) and a C/O ratio of 0.55 for a 1 $M_\mathrm{J}$ planet. The intrinsic luminosity is set by the scaling law from \citet{Thorngren_2018}. The core mass fraction is fixed to 0.15.}
    \label{fig:deviation_solarCO_1Mj}
\end{figure*}

In both atmospheric boundary conditions and Rosseland mean opacities, it is common to modify only the metallicity, while assuming that other elemental ratios, for instance the C/O ratio, are solar. However, we have shown that, in some cases, the C/O ratio does have a significant effect on the pressure and temperature at the atmospheric boundary and therefore also on the computed radius. In Fig.~\ref{fig:effect_of_co_oneplanet}, the radius difference relative to a solar C/O ratio ranges from 1 - 34 per cent for a C/O ratio of 0.2 and from 1 - 41 per cent for a ratio of 0.95, depending on the intrinsic luminosity. 

Apart from intrinsic luminosity, this percentage also depends on equilibrium temperature, atmospheric metallicity and surface gravity (see Fig.~\ref{fig:effect_of_co_differentvariations}). Therefore, we aim to determine the exact parameter space where this radius difference is significant and exceeds observational uncertainties. In this experiment, we fix the intrinsic luminosity by adopting \cite{Thorngren_2018}'s scaling relations of intrinsic heating with equilibrium temperature.

Fig.~\ref{fig:deviation_solarCO_1Mj} shows, for different equilibrium temperatures and atmospheric metallicities, the radius difference between a solar C/O ratio and a C/O ratio of 0.95 and 0.2 for a 1 $M_\mathrm{J}$ planet with a core mass fraction of 0.15. In Appendix~\ref{ap:deviationradius_solarCO}, we show similar results for a 0.3 and a 5 $M_\mathrm{J}$ planet. For a C/O ratio of 0.95, the atmospheric opacities are close to a minimum, because of the lack of dominant opacity carriers. Therefore, the radius is almost always smaller than the radius obtained for a solar C/O ratio. This trend only does not hold for equilibrium temperatures of 1000 K. There, major opacity carriers may be more abundant, because the temperature in the atmosphere is not high enough to favour the conversion from CH$_4$ to CO. For a C/O ratio of 0.2, the opacities are at a maximum due to the opacity of the very abundant H$_2$O. As a result, the radius is almost always larger than for a solar C/O ratio. Again, this trend does not hold for an equilibrium temperature of 1000 K. 

For both a C/O ratio of 0.95 and a C/O ratio of 0.2, the differences in radius peak at an equilibrium temperature of 2000 K and an atmospheric metallicity of 10 $\times$ solar. The peak in equilibrium temperature is the result of a balance between the opacity effect becoming more important for higher equilibrium temperatures at fixed intrinsic luminosities (see Fig.~\ref{fig:effect_of_co_differentvariations}), and the intrinsic luminosity peaking at an equilibrium temperature of approximately 1750\,K \citep{Thorngren_2018}. The peak at a metallicity of 10 - 30 $\times$ solar follows from the radius trend with metallicity found in Fig.~\ref{fig:effect_of_metallicity_oneplanet}. Higher radii correspond to lower surface gravities and larger scale heights, leading to more pronounced differences in radius. 

Current observational radius uncertainties depend on the exact target and star, but generally vary between $\sim$1-5 per cent for hot Jupiters (see for example these TESS surveys: \citet{Yee_2022, Yee_2023, Yee_2025, MacDougall_2023}). To identify the parameter space where the radius difference from varying C/O is significant, we set a limit in the middle of this range; at 3 per cent. For both (extreme) C/O ratios, this limit is exceeded for equilibrium temperatures $\gtrsim$ 1500\,K and atmospheric metallicities $\gtrsim$ 1 $\times$ solar (0.0 dex). For planets that fall in this regime, the assumption of a solar C/O ratio may therefore lead to erroneous results. With upcoming exoplanet missions, such as PLATO \citep{Rauer_2025}, radius uncertainties will continue to improve, potentially dropping below 1\% for ideal targets. As this precision improves, we expect this regime to extend to lower atmospheric metallicities as well.

\section{Case-study: WASP-19b} \label{sec:WASP19b_retrieval}

\begin{table*}
    \centering
    \caption{Observational constraints and references used to retrieve the interior properties of WASP-19b.}
    \begin{tabular}{l|c|c}
    \hline
     Parameter                              &  Value            & Reference                                       \\ \hline
     $M_\mathrm{planet}$ ($M_\mathrm{J}$)   &  $1.154\pm0.08$   & \cite{cortes-zuleta_2020}  \\
     $R_\mathrm{planet}$ ($R_\mathrm{J}$)   &  $1.415\pm0.046$  & \cite{cortes-zuleta_2020}  \\ 
     $k_{22}$                               &  $0.20\pm0.025$   & \cite{Bernabo_2024}          \\ 
     $T_\mathrm{day}$ (K)                   &  $2494\pm69$      & \cite{Tumborang_2024}     \\ 
     $[M/H]$   (eq. chem. / free chem.)      &  $-0.309\pm0.017$  / $-0.225\pm0.1$   &  \cite{Saha_2026}   \\ 
     $C/O$     (eq. chem. / free chem.)      &  $0.897\pm0.003$  /  $0.77\pm0.06$    &  \cite{Saha_2026}   \\ \hline
    \end{tabular}
    \label{tab:observations_WASP19b}
\end{table*}

To investigate the effect of atmospheric C/O ratio for a real planet, we retrieve the interior properties of WASP-19b under different assumed C/O ratios. Due to the inflation problem \citep{Guillot_2002, Demory_2011, Miller_2011}, which causes a degeneracy between bulk metallicity and intrinsic heat, it can be challenging to constrain the interior properties of hot Jupiters. This degeneracy can be reduced by obtaining an additional constraint on the interior, such as the Love number, which is a measure of the mass concentration in the planetary interior. The Love number of WASP-19\,b has been determined through the apsidal precession of its orbit \citep{Bernabo_2024}, making the planet a particularly useful target for this study.

Our calculation of the Love number (see Section\,\ref{sec:coupling_interior_atmosphere}) assumes a slowly rotating planet and a linear response to the perturbation. WASP-19~b has a rotational parameter of $q_0 \sim 0.06$, which is above the limit identified by \citet{Wahl_2021} below which nonlinear contributions to the response can safely be neglected. Consequently, the Love number calculated with our method may not fully capture the rotational effects on the response of WASP-19~b. A more accurate treatment of these effects is beyond the scope of this work, which focuses on the influence of atmospheric parameters on the inferred interior properties rather than on the detailed calculation of the Love number itself. We therefore use WASP-19\,b primarily as an illustrative example of the impact of the atmospheric C/O ratio, and caution that the resulting interior parameters should not be interpreted as precise measurements of the planet's true interior structure.

In Fig.~\ref{fig:deviation_solarCO_1Mj}, WASP-19\,b is located between equilibrium temperatures of 2000 and 2500 K and atmospheric metallicities of -0.5 and 0.0 dex. This region is right on the border of what we expect to result in a significant difference in derived interior properties based on the observational uncertainty in the radius only, which is 3.2 per cent for this planet.

\subsection{Retrieval framework}

We use PyMultinest \citep{Feroz_2008, Buchner_2014}, a Nested Sampling algorithm, to retrieve the interior properties. We assume normally distributed uncertainties for parameters that we fit for: the planetary radius and Love number. We set a uniform prior for $m_\mathrm{core}$ between 0 and 0.5, a log-uniform prior for $L_\mathrm{int}$ between $10^{24}$ and $10^{29}$ erg/s and set Gaussian priors on $M_\mathrm{planet}$, $T_\mathrm{eq}$, $[M/H]$ and $C/O$, centred around the observational values. The observational constraints, including references, are shown in Table\,\ref{tab:observations_WASP19b}. 

The atmospheric constraints are derived from \cite{Saha_2026}, who studied the planet's emission spectrum using JWST observations. They retrieved molecular abundances of several species in the atmosphere assuming either equilibrium chemistry or free chemistry. However, the resulting metallicity and C/O ratio are only reported for the free-chemistry case. For a consistent comparison, we compute the metallicity and C/O ratio directly from the retrieved molecular abundances ourselves for both cases. 

To obtain a metallicity and C/O ratio from the molecular abundances, we fit an equilibrium chemistry model \citep[\texttt{easyCHEM}]{Molliere_2017_easychem, Lei_2025_easychem} to the molecular abundances. This method is explained in more detail in \cite{Lothringer_2026}. We assume solar abundances from \cite{Asplund_2021} and fit a metallicity and C/O ratio at a temperature of 2425 K at 0.1\,bar. The chosen pressure falls within the range of pressures probed in the analysis of \cite{Saha_2026}, and the temperature is visually obtained from their fig. 3. The resulting metallicity and C/O ratio for both the equilibrium chemistry and free chemistry approach are shown in Table\,\ref{tab:observations_WASP19b}. 

At the high dayside temperatures considered here ($T_\mathrm{day}\sim$2500\,K), chemical timescales are expected to be short compared to transport timescales and thermochemical equilibrium provides a good approximation for the atmospheric composition \citep[e.g.][]{Moses_2011}. Therefore, we primarily show the results that use the atmospheric metallicity and C/O ratio obtained with the equilibrium chemistry retrieval as prior information. The free-chemistry results are shown in Appendix\,\ref{ap:freechem_WASP19b}.

The incoming stellar flux in our atmospheric models is equally distributed over the planet (see Section\,\ref{sec:atmosphere}). However, since emission spectra probe the dayside of hot Jupiters, a dayside-averaged atmospheric structure provides a more appropriate approximation. To account for this discrepancy, we adopt the derived dayside temperature as a proxy for the equilibrium temperature, which results in a pressure-temperature profile similar to that of a planet with flux distributed only over its dayside. 

\subsection{Interior properties for different C/O ratios}

\begin{figure*}
    \centering
    \includegraphics[width=1.7\columnwidth]{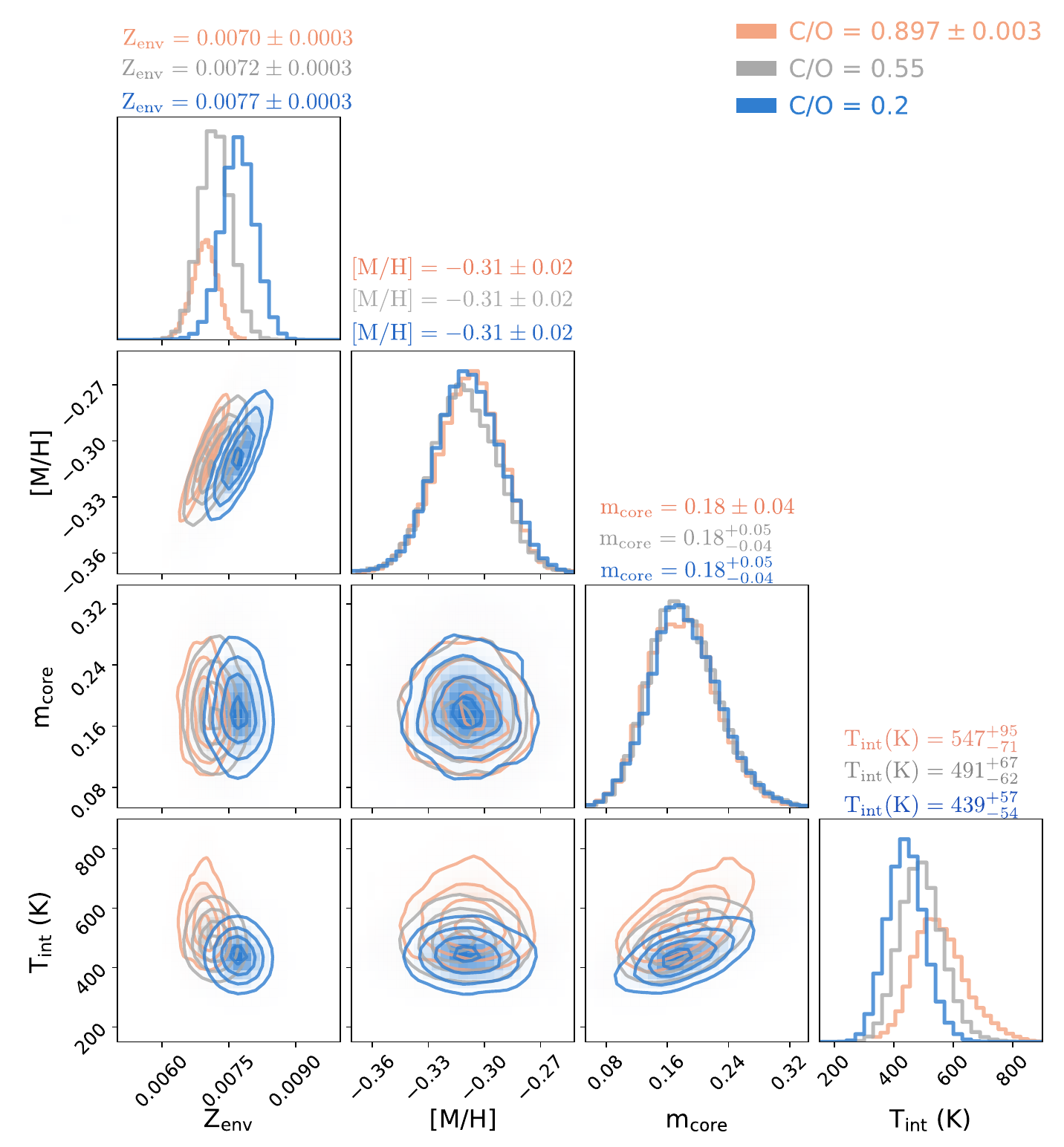}
    \caption{Retrieved envelope metal mass fraction ($Z_\mathrm{env}$), atmospheric metallicity ([M/H]), core mass fraction (m$_\mathrm{core}$) and intrinsic temperature (T$_\mathrm{int}$) for WASP-19b for different C/O ratios. }
    \label{fig:cornerplot_WASP19b_differentCO}
\end{figure*}

To compare retrieved interior properties for different atmospheric C/O ratios, we run three different retrievals: one retrieval with the observed (eq. chem) C/O ratio from \cite{Saha_2026} and two with fixed C/O ratios of 0.2 (lowest possible value in the grid) and 0.55 (solar). Fig.~\ref{fig:cornerplot_WASP19b_differentCO} shows the posterior probability distributions of the inferred interior properties.

The properties that vary the most with different C/O ratios are the intrinsic temperature ($T_\mathrm{int}$) and the envelope metallicity ($Z_\mathrm{env}$). For higher C/O ratio, we find higher and less constrained intrinsic temperatures.  This result follows naturally from the synthetic planet simulations in Fig.~\ref{fig:effect_of_co_oneplanet}. Because the radius becomes smaller with higher C/O ratio for a fixed intrinsic luminosity (in the C/O<1 regime), lower intrinsic luminosities are necessary to fit the radius for a low C/O ratio and higher intrinsic luminosities for a high C/O ratio. For a C/O ratio close to 1, the radius variation with intrinsic luminosity is the smallest, which explains the higher uncertainty in inferred intrinsic temperatures for higher C/O ratio. 

Secondly, there is some variation in the retrieved envelope metallicities for different C/O ratios. For the same atmospheric metallicity, but higher C/O ratio, we derive slightly lower envelope metallicities, similar to the $Z_\mathrm{env}$ trend observed in Fig.~\ref{fig:effect_of_co_oneplanet}. The difference is only of the order $\sim$0.001 and is present because the atmospheric metallicity constraint is very precise. This difference disappears when using the more conservative free retrieval estimates for the atmospheric metallicity and C/O ratio (only used for the orange retrieval), as shown in Appendix\,\ref{ap:freechem_WASP19b}.  

The derived core mass fraction is similar between different retrievals, because it is mostly determined by the Love number. The Love number is sensitive to the mass concentration and thus will be primarily affected by the core mass fraction and not the atmospheric composition, which mostly change the overall temperatures and densities throughout the planet. Because the inferred core masses are similar and the core dominates the planet’s metal budget (with a metallicity of 0.18, compared to 0.007 in the envelope), the retrieved bulk metallicity is also very similar across different C/O ratios.

\subsection{Implications for other planets}
These results imply that the assumption of a solar C/O ratio may not be valid for interior structure retrievals, particularly for hot Jupiters. Our case-study, WASP-19\,b, is situated in a parameter space where different C/O ratios start to affect interior retrievals. For this planet, C/O ratios closer to 1 lead to higher retrieved intrinsic temperatures than lower C/O ratios. The intrinsic temperature is derived without assuming any physical inflation mechanism, which is possible because of the precise constraints on the Love number and the atmospheric properties \citep{vandijk_2025}. To date, WASP-19\,b is the only planet with both a constrained Love number and a well-characterised atmosphere. However, the bulk metallicities of other planets can be determined by assuming a population-derived intrinsic heating that depends on stellar irradiation \cite[e.g.][]{Thorngren_2018, Sarkis_2021,Schmidt_2026}. If a solar C/O ratio is assumed, these bulk metallicities may deviate substantially from those obtained using the observed C/O ratio. These implications apply to systems with high equilibrium temperatures, high atmospheric metallicities and C/O ratios that deviate substantially from solar, such as HD 149026\,b (T$_\mathrm{eq}$ = 1694 K, [M/H] = 1.60, C/O = 0.84), WASP-121\,b (T$_\mathrm{eq}$ = 2449 K, [M/H] = 0.78, C/O = 0.89) and WASP-33\,b (T$_\mathrm{eq}$ = 2784 K, [M/H] = 0.68, C/O = 0.8) \citep{Lothringer_2026}.

\section{Discussion}

\subsection{Sensitivity to the assumed transit radius pressure} \label{sec:transit_radius_definition}

\begin{figure}
    \centering
    \includegraphics[width=\columnwidth]{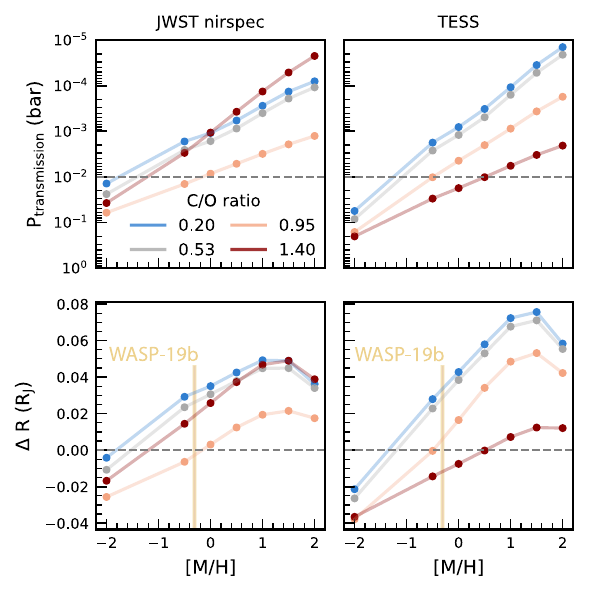}
    \caption{The transmission pressure (top) and the corresponding difference in radius compared to the assumed transmission pressure (bottom) for different instrumental throughput functions and different atmospheric compositions. Different colours correspond to different C/O ratios. WASP-19b's observational uncertainties on the radius and atmospheric metallicity are shown for reference. }
    \label{fig:transmission_radius}
\end{figure}

In atmospheric boundary grids, the transit radius is typically assumed to be at a fixed pressure level. In this work, it is fixed at 10 mbar. In reality, the transit radius pressure depends on wavelength and atmospheric opacities. In this section, we compare this assumed pressure with a more accurately calculated transit radius pressure. The transit pressure or radius for a specific wavelength can be determined by calculating the transmission of stellar light through the atmosphere of the planet at a grazing angle (see Section\,2.2.1 in \cite{Molliere_2019}). Repeating this calculation over all wavelengths yields the transmission spectrum. The transmission radius and corresponding pressure for a specific atmosphere and telescope observation can be computed (see equation\,\ref{eq:transmission_radius}) from this transmission spectrum $R_{p} (\lambda)$ by integrating over the instrument's wavelength-dependent throughput function $\mathcal{T}(\lambda)$, which describes the combined transmission and detection efficiency of the telescope, instrument and detector as a function of wavelength.

\begin{equation} \label{eq:transmission_radius}
    R_{p, {\rm transmission}} = \int_\lambda \mathcal{T}(\lambda) R_p(\lambda) d\lambda
\end{equation}

Fig.~\ref{fig:transmission_radius} shows the resulting transit radius pressure (top panel) and the difference between the corresponding transit radius and the radius corresponding to the assumed pressure of 10 mbar (bottom panel) for different atmospheric compositions and two different instruments (JWST NIRSpec G395H/F290LP (2.87-5.27 micron), \citet{Giardino_2022, Pandeia_calculator}, and TESS (0.6-1.0 micron), \citet{Ricker_2015}). Other atmospheric grid parameters are fixed to values close to the dayside atmosphere of WASP-19\,b, with surface gravity and effective temperature set to 1000\,cm\,s$^{-2}$ and 2500\,K, respectively. 

For both instruments, higher atmospheric metallicity corresponds to lower transmission pressures, because of the increase in opacity with higher metallicity. As a result, the radius difference increases with higher atmospheric metallicity until a metallicity of $\sim$ 10-30 $\times$ solar. For even higher metallicities, the transmission pressure continues to decrease; however, the accompanying increase in the mean molecular weight reduces the atmospheric scale height, so a given decrease in pressure translates into a smaller increase in radius than at lower metallicities.

For different C/O ratios (different colours in Fig.~\ref{fig:transmission_radius}), the trend depends on the instrument. In the JWST NIRspec band (2.87 - 5.27 micron), opacity of species, such as H$_2$O, CH$_4$, HCN and C$_2$H$_2$, dominates. Therefore, there is again this opposing effect between C/O > 1 and C/O < 1, where the opacity decreases with higher C/O (due to decreasing abundance of H$_2$O), until C/O$\approx$1, after which the opacity increases again (due to increasing abundances of CH$_4$, HCN and C$_2$H$_2$). With higher opacity, the transmission pressure accordingly decreases, resulting in a larger radius increase for either low C/O ratio (0.2) or high C/O ratio (1.4). In the TESS band (0.5 - 1.0 micron), TiO and VO are the dominant opacity species. Therefore, as TiO and VO become less abundant with higher C/O, the transmission pressure simply increases with higher C/O ratio. 

Because the observed transit radius depends on the instrument and the atmosphere, our results could change when taking this dependency into account. Replacing our assumed transit radius pressure of 10 mbar with the calculated transmission pressure would only marginally affect the radius trend with C/O ratio presented in Section\,\ref{sec:effect_of_coratio}. The exact impact depends on the observational bandpass. In the JWST NIRSpec band, the trend would be slightly amplified, as accounting for the more accurately calculated transmission pressure leads to larger radii at C/O=0.2 and C/O=1.4 relative to the radius minimum at C/O=0.95. In contrast, in the TESS band, where TiO and VO dominate the opacity, the radius increase at C/O=1.4 is smaller than at C/O=0.95. Therefore, the rise in radius observed for C/O>1 (in Section\,\ref{sec:effect_of_coratio}) may be reduced or even disappear when observations are made in a bandpass similar to that of TESS. 

For the atmosphere of WASP-19\,b (highlighted in yellow in Fig.~\ref{fig:transmission_radius}), the radius difference (lower panel) of maximally $\sim0.043$\,$R_{\rm J}$ is below the 1-sigma observational uncertainty of 0.046\,$R_{\rm J}$ on the radius, indicating that our assumption of a constant transit radius pressure at 10 mbar suffices for this planet. However, for higher atmospheric metallicities, the transit radius pressure moves to higher altitudes, up to approximately 0.1-0.01\,mbar for metallicities of 10-100\,$\times$\,solar, which can lead to an increase in radius up to 0.08\,R$_\mathrm{J}$. For a radius similar to WASP-19\,b, this increase corresponds to 5 per cent of the radius, which exceeds typical observational uncertainties. For planets with a lower surface gravity the radius variation is even larger, reaching a maximum of 0.4\,R$_\mathrm{J}$ for the lowest surface gravity in our grid of 200\,cm\,s$^{-2}$. Hence, the transit radius should be computed more carefully for planets with higher atmospheric metallicities or lower surface gravities, as the assumption of a fixed transit radius pressure may introduce significant errors. Some examples of planets that fit in this category include WASP-17\,b \citep[$g$ = 320\,cm\,s$^{-2}$ and super-solar metallicity,][]{Bonomo_2017, Gressier_2025}, WASP-107\,b \citep[$g$ = 269\,cm\,s$^{-2}$, metallicity of 17 $\times$ solar,][]{Howard_WASP107b_2025, Huang_2026} and WASP-127\,b \citep[$g$ = 238\,cm\,s$^{-2}$, super-solar metallicity,][]{Seidel_2020, Kanumalla_2024}.

\subsection{Limitations}

Atmospheric grids are limited by multifunctionality and computational time. In this paper, we focus on the effect of atmospheric composition on the interiors of hot Jupiters. Therefore, we limit other dimensions of the grid by applying several simplifying assumptions that keep computational time feasible and the grid multifunctional over a wide parameter space.

Firstly, we assume equilibrium chemistry in all atmosphere models. In reality, disequilibrium processes, such as photochemistry and atmospheric dynamics, may alter abundances of species and, consequently, the pressure-temperature structure. Photochemistry is an important process for hot Jupiters because of the high levels of stellar irradiation \citep[e.g.][]{Tsai_2021, Line_2010, Moses_2011, Shulyak_2020, Venot_2012, Zahnle_2009}. Nonetheless, \cite{Drummond_2016} show that this process only becomes important at pressures below $\sim10^{-5}$ bar, where the optical depth is low, and produces minor changes to the temperature structure deeper in the atmosphere. While photochemistry could in principle affect deeper temperatures indirectly through changes in atmospheric opacity and redistribution of absorbed stellar radiation, such effects appear to be small for the hot Jupiter conditions and atmospheric species explored by \cite{Drummond_2016}. Vertical mixing is expected to have a more prominent effect on the deep atmospheric temperature profile. \cite{Drummond_2016} show that, when this process is important, it can lead to deviations in the pressure-temperature profile up to 100 K. However, at high atmospheric temperatures, chemical timescales become shorter than chemical timescales and disequilibrium processes gradually become less important. As a result,  these processes can be safely ignored at sufficiently high photospheric temperatures \citep[T $\gtrsim$ 2000 K,][]{Moses_2014}. The atmospheric grid should be used with caution for planets on the low temperature end. 

Secondly, for individual planets, certain opacity sources might be missing. In particular, clouds can be a significant source of continuum opacity and therefore influence the atmospheric pressure-temperature profile. Although, at sufficiently high temperatures, condensation is not expected, it can occur at the lower equilibrium temperature end of hot Jupiters. In fact, observations have repeatedly revealed clouds in the atmospheres of hot Jupiters \citep[e.g.][]{Sing_2016, Inglis_2024, Grant_2023}. For the specific case of WASP-19\,b, \cite{Saha_2026} detected a hint of SiO$_2$ clouds on the dayside. However, if confirmed, these clouds likely are present in upper layers of the atmosphere, around the mbar level. Therefore, these clouds are expected to primarily affect the transmission level of the atmosphere through their additional opacity, potentially altering the inferred planetary radius. Any effects of clouds on the atmospheric structure, including possible indirect changes to the deeper thermal profile, are neglected in the present atmospheric grid.

For planets with detected clouds or planets with lower equilibrium temperatures, where clouds likely exist, cloud opacity could change the atmospheric temperature-profile. The exact effect on the temperature-pressure profile depends on the exact properties of the cloud opacity. \cite{Heng_2012} showed that a purely absorbing cloud cools the upper atmosphere, while it heats the lower atmosphere, whereas the inclusion of short-wavelength scattering produces the opposite effect. \cite{Goyal_2020} showed that the treatment of condensation (rainout or local condensation) can also alter the pressure-temperature profile, resulting in a hotter deep atmosphere by approximately 200 K with the rainout-approach. 

Despite these model-dependent uncertainties in the treatment of clouds, the general impact of clouds on the interiors of gas giant planets has been investigated in several studies \cite[e.g.][]{Poser_2024, Siebenaler_2026}. These studies agree that clouds delay the cooling and contraction of the planet during evolution. However, because cloud physics in these studies is treated in a simplified manner, further work is needed to fully characterize their effect on interiors and evolution. Since the appropriate cloud treatment depends on the specific planet under consideration, a detailed treatment of clouds is beyond the scope of the current study, which relies on a pre-computed atmospheric grid rather than planet-specific atmospheric models.

\section{Conclusions} 

In this study, we investigate how atmospheric composition affects, through the atmospheric boundary condition, the interiors of hot Jupiters. To this end, we construct a grid of self-consistent atmosphere models, allowing for variations in atmospheric composition through both the metallicity and the C/O ratio. Aside from that, the grid varies in intrinsic temperature, effective temperature and surface gravity. This grid is coupled to a static interior structure code (\texttt{CEPAM}) at the RCB. We apply the framework to a range of synthetic planets to isolate the effect of atmospheric metallicity and C/O ratio on the atmospheric boundary, interior and radius of the planet. Additionally, to demonstrate the impact of C/O ratio on inferred properties, we retrieve the interior properties of WASP-19~b assuming a set of different C/O ratios.

Our results show that both metallicity and C/O ratio significantly alter the pressure and temperature at the RCB. Increasing metallicity enhances the opacity, shifting the RCB to lower pressures and raising  temperatures through the interior of the planet, which ultimately inflates the planet. However, at metallicities above $\sim 10\times$\,solar, the planet contracts again, due to the increase in mean molecular weight. Under fixed metallicity, the C/O ratio determines the dominant molecular species in the atmosphere. Therefore, as the C/O ratio changes, the opacity varies, resulting in different pressures and temperatures at the RCB. For our 1\,$M_\mathrm{J}$ synthetic planet, C/O ratio differences can change the temperature at the RCB up to $\sim 300$\,K. This temperature change propagates through the interior and results in radius differences between 1-41 per cent, depending on the intrinsic luminosity and C/O ratio.

These results demonstrate that commonly adopted assumptions, such as a solar C/O ratio, in the atmospheric boundary condition used in interior models can introduce systematic biases in interior structure inferences. In the era of JWST, where atmospheric abundances are increasingly constrained observationally, such simplifications are no longer justified. We find that for hot Jupiters with equilibrium temperatures $\gtrsim 1500$\,K and super-solar metallicities, assuming a solar C/O ratio can lead to radius discrepancies that exceed current observational uncertainties ($\sim 3$ per cent), reaching up to 27 per cent for a synthetic 1~M$_\mathrm{J}$ planet with a core mass fraction of 0.15. This has direct consequences for inferred intrinsic temperatures, as illustrated for our case-study WASP-19~b, or inferred bulk metallicities. Upcoming exoplanet missions, such as PLATO, will improve observational uncertainties, extending the impacted planetary regime to even lower atmospheric metallicities. 

As observational constraints on atmospheric composition continue to improve, interior models should be updated accordingly. For highly irradiated planets such as hot Jupiters, the extended atmosphere makes the choice of atmospheric boundary condition a key source of systematic uncertainty, and one that must be treated with increasing physical realism.

\section*{Acknowledgements}
This work has been done thanks to support from the European Research Council (ERC) under the European Union’s Horizon 2020 research and innovation programme (grant agreement no. 101088557, N-GINE). We acknowledge the use of AI-assistend tools, including GitHub Copilot for code completion and AI-based language tools (ChatGPT and Grammarly) for proofreading and improving the clarity of the manuscript. All outputs were reviewed and verified by the authors.

\section*{Data Availability}

The grid of atmospheric boundary condition profiles generated for this work is available at \url{https://doi.org/10.5281/zenodo.20629944}.

\bibliographystyle{mnras}
\bibliography{extraref.bib}

\appendix

\section{Comparison between metallicity conversions} \label{ap:metallicity_conversion}

\begin{figure}
    \centering
    \includegraphics[width=\columnwidth]{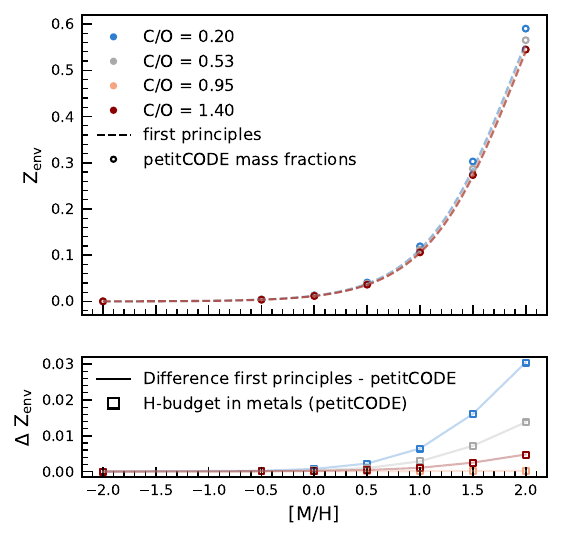}
    \caption{Metal mass fractions at the RCB calculated with petitCODE as a function of log-metal enrichment (compared to solar). These results are compared with a metal mass fraction calculation from first principles, i.e. assuming a solar elemental abundance set (see Appendix A from \citet{Fortney_2013}). The bottom panel shows the difference in metal mass fraction between chemical equilibrium calculations (petitCODE) and first principles, as well as the calculated mass-budget of hydrogen locked up in (under metals categorized) molecules in petitCODE.}
    \label{fig:metallicity_conversions}
\end{figure}

In our coupled interior-atmosphere model, we set the envelope metal mass fraction ($Z_{\rm env}$) in the interior equal to the metal mass fraction at the RCB calculated with petitCODE (computed assuming chemical equilibrium). In this appendix, we compare the resulting metal mass fractions with mass fractions calculated through first principles, i.e. only assuming a (modified) solar elemental abundance set. This metal mass fraction can be expressed as (see e.g. \citet{Fortney_2013} for a derivation)

\begin{equation} \label{eq:metal_conversion}
    Z = \frac{\sum_i \mu_i N_i/N_{\rm H}}{\mu_{\rm H} + \mu_{\rm He} N_{\rm He}/N_{\rm H} + \sum_i \mu_i N_i/N_{\rm H}}, 
\end{equation}

with $\mu_i$ the atomic weight of species $i$ and $N_i/N_{\rm H}$ the relative abundance of species $i$ compared to hydrogen. We calculate the metal mass fraction for a specific metallicity and C/O ratio by modifying a solar abundance set \citep{Asplund_2021} according to our adopted definitions of metallicity and C/O ratio (see Section\,\ref{sec:atmosphere}) and plugging the resulting relative abundance ratios into equation\,\ref{eq:metal_conversion}. Fig.\,\ref{fig:metallicity_conversions} shows these first-principles computed mass fractions and the mass fractions calculated with petitCODE for an effective temperature of 2500 K, an internal temperature of 550 K and a surface gravity of 1000\,cm\,s$^{-2}$. The bottom panel shows the difference in metal mass fraction between the two calculations, which can reach up to 0.03 for a C/O ratio of 0.2 and a metallicity of 100\,$\times$\,solar. The main difference between these calculations is that the first-principles calculations assumes a purely atomic mixture, while in chemical equilibrium calculations (at hot Jupiter atmospheric temperatures) these atoms are locked up in molecules. Therefore, in chemical equilibrium calculations, some of the hydrogen-budget is locked up in molecules that are categorized under metals (most importantly H$_2$O, CH$_4$ and C$_2$H$_2$). We show the mass fraction of hydrogen atoms that are locked up in (under metal categorized) molecules in the bottom panel of Fig.\,\ref{fig:metallicity_conversions}, explaining the differences between first-principles and chemical equilibrium calculation of the metal mass fraction. For a C/O ratio of 0.95, the difference is small, because the most abundant species in the atmosphere is CO and hydrogen-containing species have lower abundances. But for either C/O<0.95, where H$_2$O becomes more abundant, or C/O>0.95, where CH$_4$ and C$_2$H$_2$ become more abundant, the hydrogen mass budget in metals increases.

\section{Radius difference compared to solar C/O ratio for different masses} \label{ap:deviationradius_solarCO}

\begin{figure*}
    \centering
    \includegraphics[width=1.9\columnwidth]{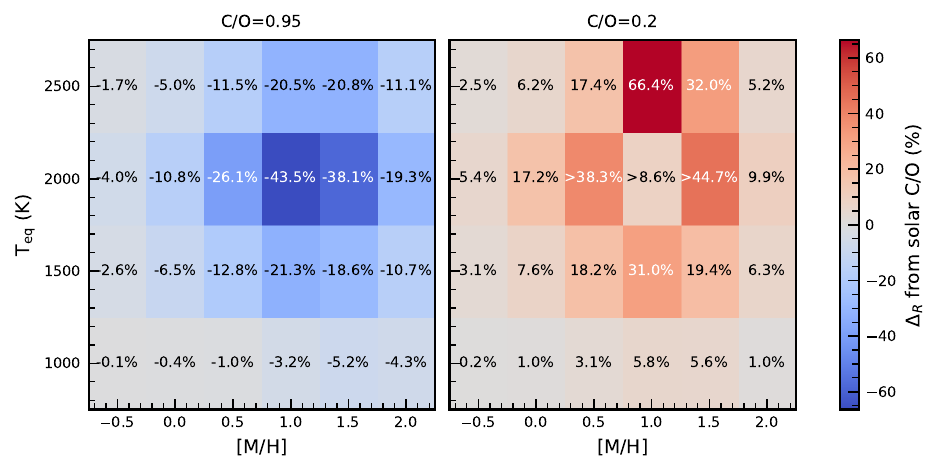}
    \caption{Radius difference between a C/O ratio of 0.95 (left) / 0.2 (right) and a C/O ratio of 0.55 for a 0.3 $M_\mathrm{J}$ planet. The intrinsic luminosity is set by the scaling law from \citet{Thorngren_2018}. The core mass is fixed to 64 M$_\oplus$. }
    \label{fig:radius_difference_compared_to_solar_0.3Mjup}
\end{figure*}

\begin{figure*}
    \centering
    \includegraphics[width=1.9\columnwidth]{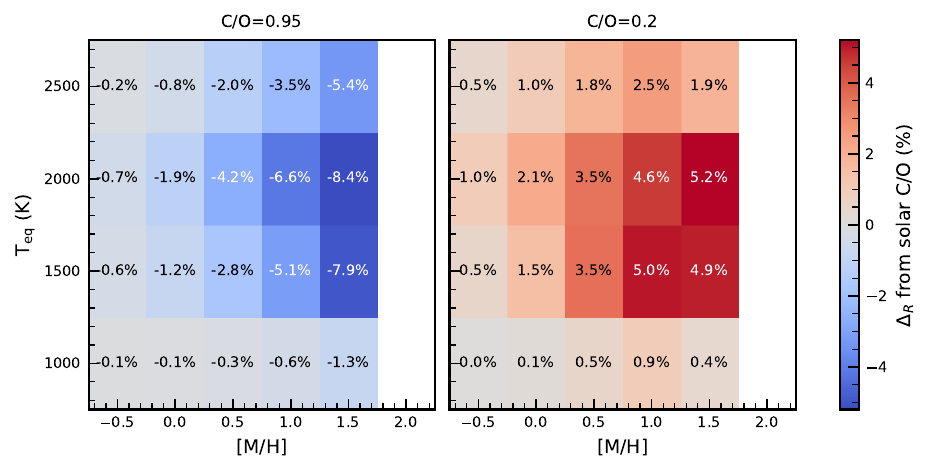}
    \caption{Radius difference between a C/O ratio of 0.95 (left) / 0.2 (right) and a C/O ratio of 0.55 for a 5 $M_\mathrm{J}$ planet. The intrinsic luminosity is set by the scaling law from \citet{Thorngren_2018}. The core mass is fixed to 30 M$_\oplus$.}
    \label{fig:radius_difference_compared_to_solar_5Mjup}
\end{figure*}

In Section\,\ref{sec:comparison_solarCO}, we present radius differences for a 1\,M$_{\rm J}$ planet between atmospheres with a solar C/O ratio and C/O ratios of 0.2 and 0.95. To illustrate how these results depend on surface gravity, we show the same analysis for a 0.3\,M$_{\rm J}$ planet (Fig.~\ref{fig:radius_difference_compared_to_solar_0.3Mjup}) with a core mass of 64\,M$_\oplus$, and for a 5.0\,M$_{\rm J}$ planet (Fig.~\ref{fig:radius_difference_compared_to_solar_5Mjup}) with a core mass of 30~M$_\oplus$.

For the 0.3\,M$_{\rm J}$ planet, we adopt a relatively large core mass because, given the lower surface-gravity limit of our grid, smaller core masses would result in radii that exceed the grid limits for most metallicities and equilibrium temperatures. Even with the adopted core mass, two interior models with a C/O ratio of 0.2 yield radii larger than the maximum radius corresponding to the minimum surface gravity of 200 cm s$^{-2}$. For these models, we therefore report lower limits on the radius differences; the actual differences are larger.

For the 5.0\,M$_{\rm J}$ planet, interior models with the highest atmospheric metallicities require envelope metallicities that are too high to reproduce the planetary mass and are therefore omitted.

The general trends for both planets are consistent with those found for the 1\,M$_{\rm J}$ case discussed in Section\,\ref{sec:comparison_solarCO}. However, the relative radius differences are generally larger for the 0.3\,M$_{\rm J}$ planet and smaller for the 5.0\,M$_{\rm J}$ planet, reflecting their lower and higher surface gravities, respectively.

\section{Free chemistry results WASP-19b} \label{ap:freechem_WASP19b}

In this appendix, we show the retrieved interior parameters of WASP-19\,b for different C/O ratios, using the free chemistry results of \cite{Saha_2026} to constrain the atmospheric metallicity and C/O ratio (see Table\,\ref{tab:observations_WASP19b}). The inferred interior parameters are similar to those obtained in Section\,\ref{sec:WASP19b_retrieval}. However, differences in inferred envelope metal mass fractions ($Z_\mathrm{env}$) for different C/O ratios are now not apparent, because of the larger uncertainty in the atmospheric metallicity constraint. 

\begin{figure*}
    \centering
    \includegraphics[width=1.7\columnwidth]{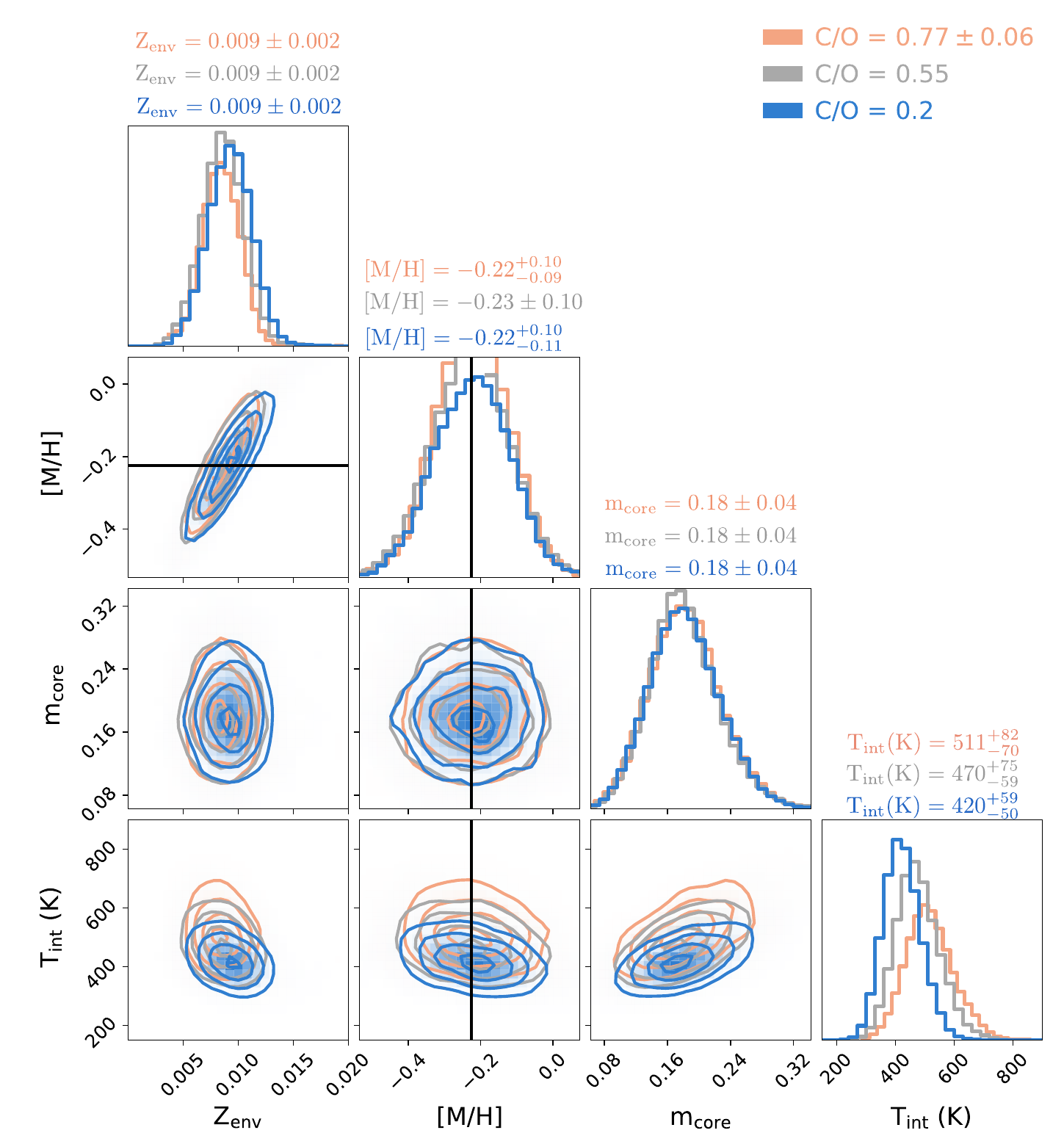}
    \caption{Retrieved envelope metal mass fraction ($Z_\mathrm{env}$), atmospheric metallicity ([M/H]), core mass fraction (m$_\mathrm{core}$) and intrinsic temperature (T$_\mathrm{int}$) for WASP-19b with different C/O ratios. For the atmospheric metallicity and C/O ratio (only used in the orange retrieval), we adopt the constraints resulting from \citet{Saha_2026}'s free-chemistry approach.}
    \label{fig:placeholder}
\end{figure*}

\bsp	%
\label{lastpage}
\end{document}